\documentclass[12pt]{article}

\usepackage{amssymb,amsmath,mathrsfs,tikz,graphicx}
\newcommand{\bm}[1]{\mbox{\boldmath $#1$}}

\def\S{\Sigma}

\def\Aplus{k-\frac{R_+}{20} a^2}
\def\Aminus{k-\frac{R_-}{20} a^2}
\def\A{k-\frac{R_\pm}{20} a^2}

\def\B1{\dot a^2 +\Aplus}
\def\B2{\dot a^2 +\Aminus}
\def\B{\dot a^2 +\A}

\def\be{\begin{equation}}
\def\ee{\end{equation}}
\def\bea{\begin{eqnarray}}
\def\eea{\end{eqnarray}}
\def\bean{\begin{eqnarray*}}
\def\eean{\end{eqnarray*}}

\begin{document}

\title{Gravitational double layers 
}
\author{Jos\'e M. M. Senovilla \\
F\'{\i}sica Te\'orica, Universidad del Pa\'{\i}s Vasco, \\
Apartado 644, 48080 Bilbao, Spain \\ 
josemm.senovilla@ehu.es}
\date{}
\maketitle
\begin{abstract} 
I analyze the properties of thin shells through which the scalar curvature $R$ is discontinuous in gravity theories with Lagrangian $F(R)=R-2\Lambda+\alpha R^2$ on the bulk. These shells/domain walls are of a new kind because they possess, in addition to the standard energy-momentum tensor, an external energy flux vector, an external scalar pressure/tension and, most exotic of all, another energy-momentum contribution resembling classical dipole distributions on a shell: a double layer. I prove that all these contributions are necessary to make the entire energy-momentum tensor divergence-free. This is the first known occurrence of such a type of double layer in a gravity theory. I present explicit examples in constant-curvature 5-dimensional bulks, with a brief study of their properties: new physical behaviors arise.
\end{abstract} 


In classical electromagnetism a surface charge distribution creates an electric field with a jump in its component normal to the surface, while a double layer (or dipole) distribution of charges has a discontinuous electrostatic potential \cite{J}. Mathematically, the former can be appropriately described with a Dirac delta supported on the surface and has an analog in Gravitation describing matter concentrations  on a thin shell ---or a domain wall or braneworld--- \cite{I,GT}. Similarly, the latter can be mathematically described via the normal derivative of a Dirac delta supported on the surface \cite{SW,TKT}. However, it has no analog in Gravitation, and such gravitational dipole layers are assumed to be unphysical on the grounds that dipole distributions are impossible due to the positivity of masses and the attractiveness of gravity. The purpose of this paper is to show that double layers do exist in some gravitational theories, as follows from results in \cite{S}. In particular, I provide the matter content and field equations for gravitational double layers in quadratic $F(R)$ theories (e.g. \cite{SF,CF,NO}). These are the first examples of such idealizations in a gravitational theory. It must be remarked, however,  that the possibility of having effective dipolar matter, polarizable in a gravitational field, has been considered as a viable possibility to describe dark matter and energy within General Relativity (GR) in a series of interesting papers, see \cite{BL1} and references therein. On the other hand,  the pure (non-gauge) dipole contributions arising in the  ``Blackfold" formalism \cite{ACHO,CE,AGO} have a different status as they concern the equations of motion for {\em test} $p$-branes, neglecting the back-reaction of the localized object onto the bulk. 

For a quadratic theory of gravity with Lagrangian density given by
$$
F(R)=R-2\Lambda +\alpha R^2
$$
where $\Lambda$ (the bulk cosmological constant) and $\alpha$ are constants, the field equations read
\be
(1+2\alpha R)R_{\mu\nu}-\frac{1}{2}F(R)g_{\mu\nu}-2\alpha\left( \nabla_\mu \nabla_\nu R -g_{\mu\nu} \nabla_\rho\nabla^\rho R\right)=\kappa T_{\mu\nu} \label{fe}
\ee
where $T_{\mu\nu}$ is the energy-momentum tensor of the matter fields. These theories include the important Starobinsky inflationary models \cite{St} and they possess a positive mass theorem \cite{Str} and a well-defined entropy formulation \cite{JKM}. Notice the absence of quadratic terms in $\nabla R$ in (\ref{fe}). Using this fact, one can consistently describe braneworlds (or thin shells or domain walls) $\S$ across which the scalar curvature $R$ can be {\em discontinuous}, and still the field equations (\ref{fe}) are well-defined in the distributional sense \cite{S}. This is quite an exceptional case leading to an unexpected matter content on $\Sigma$ ---that divides the manifold into two parts denoted by $V^{\pm}$--- described by an energy-momentum tensor distribution of the following form (distributions are distinguished by an underline)
\be
\underline T_{\mu\nu}=T^+_{\mu\nu} \underline\theta +T^-_{\mu\nu} (1-\underline\theta)+\left(\tau_{\mu\nu}+\tau_\mu n_\nu +\tau_\nu n_\mu +\tau n_\mu n_\nu \right) \underline\delta^{\Sigma} + \underline{t}_{\mu\nu} \label{emt}
\ee
where $n_{\mu}$ is the unit normal to $\S$ (pointing towards $V^+$), $T^\pm_{\mu\nu}$ are the energy-momentum tensors on $V^{\pm}$ respectively, $\underline\theta$ is the distribution associated to the $\S$-step function 
$$
\theta =\left\{
\begin{array}{ccc}
1 &  & V^+\\
1/2 & \mbox{on} & \Sigma \\
0 &  & V^-
\end{array}\right.
$$
while $\underline{\delta}^\Sigma$ is a scalar distribution (a Dirac delta) with support on $\Sigma$ acting on any test function $Y$ as
$$
\left<\underline{\delta}^\Sigma ,Y\right> =\int_\S Y 
$$
and such that
$$
\nabla_\mu\,   \underline{\theta} = n_\mu\,  \underline{\delta}^\Sigma \, .
$$
The expressions and proposed names for the objects in (\ref{emt}) supported on $\S$ are:
\begin{enumerate}
\item the energy-momentum tensor  $\tau_{\mu\nu}$ on $\S$, given by\footnote{Standard notation for discontinuities is used, so that for any function $f$ with definite limits on both sides of $\Sigma$: $\left[f\right](p)\equiv \lim_{x\rightarrow p}f^+(x) -\lim_{x\rightarrow p} f^-(x)$, $p\in \Sigma$, $f^\pm$ being the restrictions of $f$ to the bulk sides $V^\pm$ of $\Sigma$, respectively.}
\be
\kappa \tau_{\mu\nu} =-\{1+\alpha(R^++R^-)\}[K_{\mu\nu}] 
+\alpha \left\{2bh_{\mu\nu}-[R](K^+_{\mu\nu}+ K^-_{\mu\nu})\right\}, \, \, n^\mu\tau_{\mu\nu}=0, \label{tauexc}
\ee
where $h_{\mu\nu}=g_{\mu\nu}-n_\mu n_\nu$ is the first fundamental form on $\S$ (assumed to agree on both sides), $K^\pm_{\beta\mu}=h^\rho{}_{\beta}h^\sigma_\mu \nabla^\pm_\rho n_\sigma $ its second fundamental forms from both sides, and
$b$ (called $a$ in \cite{S})  is a function on $\S$ measuring the jump on the normal derivative of $R$ across $\S$:
$$
b = n^\mu \left[\nabla_{\mu}R\right] . 
$$
An alternative, useful, expression for (\ref{tauexc}) reads
\be
\kappa \tau_{\mu\nu} =-[K_{\mu\nu}] 
+2\alpha \left(bh_{\mu\nu}-[RK_{\mu\nu}]\right). \label{tauexc1}
\ee
$\tau_{\mu\nu}$ is the only quantity usually defined in standard shells.
\item the ``external flux momentum'' $\tau_{\mu}$ with
\be
\kappa \tau _\mu =-2\alpha \overline\nabla_\mu [R], \quad \quad n^\mu \tau_\mu =0, \label{tauex}
\ee
where $\overline\nabla$ is the Levi-Civita covariant derivative in $(\S,h_{\mu\nu})$. This momentum vector measures the normal-tangent components of $\underline{T}_{\mu\nu}$ supported on $\S$, so that its timelike component describes the normal flux of energy across $\S$ on $\S$ while its spatial components measure the normal-tangential stresses.
\item the ``external pressure or tension'' $\tau$ defined as
\be 
\kappa \tau = 2\alpha [R] K^\rho{}_\rho , \label{taue} 
\ee
where $K^\rho{}_\rho$ is the trace of {\em either} $K^\pm_{\mu\nu}$ because an indispensable requirement for shells in non-linear  $F(R)$ gravity is that \cite{DSS,BD,S} 
\be
\left[K^\mu{}_\mu\right] =0 \, .\label{Kcont}
\ee
This ensures that $\underline{R}$ is associated to a (possibly discontinuous) function $R$ without a singular distributional part \cite{MS,S}. Taking the trace of (\ref{tauexc1}) one obtains a relation between $b$, $\tau$ and the trace of $\tau_{\mu\nu}$ ($n=$ dim$(\S)$):
$$
\kappa \left(\tau^{\rho}{}_{\rho}+\tau\right) = 2\alpha n b
$$
Eq.(\ref{Kcont}) implies radical differences with GR: it forbids the use of umbilical hypersurfaces $\S$ ---characterized by $K_{\mu\nu}=f h_{\mu\nu}$--- unless they are totally geodesic. It also restricts the possibility of having $\mathbb{Z}_2$ symmetric branes to the case with zero mean curvature $K^\mu{}_\mu=0$. 

The scalar $\tau$ measures the total normal pressure/tension supported on $\S$.
\item the ``double-layer energy-momentum tensor distribution'' $\underline t_{\mu\nu}$, which is defined by acting on any test function $Y$ by 
\be
\kappa \left<\underline t_{\mu\nu},Y\right> = -2\alpha \int_\Sigma [R]h_{\mu\nu}\,  n^\rho\nabla_\rho Y \, . \label{t}
\ee
\end{enumerate}
This is a symmetric ($\underline t_{\mu\nu}=\underline t_{\nu\mu}$) tensor distribution of Dirac ``delta-prime'' type, it has support on $\S$ but its product with objects intrinsic to $\S$ is not defined unless their extensions off $\S$ are known (in particular, one cannot write $\underline t_{\mu\nu}=h_{\mu\nu} \underline t$ for some scalar distribution $\underline t$, unless $n_\mu$ is extended outside $\S$). Similarly, the divergence $\overline\nabla^{\mu}\underline{t}_{\mu\nu}$ is not defined, only the bulk divergence $\nabla^{\mu}\underline{t}_{\mu\nu}$ makes mathematical sense. This $\underline{t}_{\mu\nu}$ resembles the energy-momentum content of double layer surface charge distributions, or ``dipole distributions'', with strength $2\alpha[R]h_{\mu\nu}$. Thus, the appearance of (\ref{t}) is remarkable and very surprising.  Here, it seems to represent the idealization of an abrupt change on the scalar curvature $R$ which, one has to bear in mind, acts as a source or dynamical variable in $F(R)$ gravity. It is worth noticing here that any such gravity  is equivalent to a certain scalar-tensor theory with the Brans-Dicke parameter $\omega =0$ \cite{H,TT,Ch,FT}. In our particular quadratic case this is achieved by defining a scalar field $\phi = 2\alpha R-1$, with potential $V(\phi)=2\Lambda+(\phi^{2}-2\phi-3)/(4\alpha)$ \cite{FT,TT}. 
Thus, in this alternative viewpoint the double layer describes an abrupt discontinuity in the scalar field $\phi$.

The Bianchi identities $\nabla_{[\rho}\underline{R}_{\mu\nu]\alpha}{}^{\beta}=0$ hold for the Riemann tensor distribution $\underline{R}_{\mu\nu\alpha}{}^{\beta}$ as proven in \cite{MS}, see also {\cite{L,T}. This is all that one needs to prove the divergence-free property of the lefthand side in the field equations (\ref{fe}), and therefore the righthand side must be divergence free too. This implies that (\ref{emt}) is divergence free as a tensor distribution. Hence, a long and elaborated calculation using distributions ---in which (\ref{Kcont}) is fundamental in order to be able to use a Ricci identity for tensor distributions--- leads to
\bean
\underline{\delta}^\Sigma\left\{(1+2\alpha R_\S)K_\S^{\rho\sigma}[K_{\rho\sigma}]n_\nu -2\alpha \overline\nabla^\mu\left(R_\S[K_{\mu\nu}] \right) +2\alpha (\overline\nabla_\nu b -b K^\rho{}_\rho n_\nu )\right.\\
\left. -2\alpha K_{\S\mu\nu} \overline\nabla^\mu[R]-2\alpha n_\nu \overline\nabla^\mu \overline\nabla_\mu [R]
-2\alpha[R]n^\mu R_{\S\mu\nu}+\kappa n^\mu[T_{\mu\nu}]\right\}=0
\eean
where $R_\S=(R^++R^-)/2$ is the value of the bulk scalar curvature at $\S$, and analogously for the other objects evaluated on $\S$. Projecting into normal and tangential components one easily gets
\bean
\kappa n^{\mu}n^{\nu}[T_{\mu\nu}]&=& -(1+2\alpha R_{\S})K^{\rho\sigma}_{\S}[K_{\rho\sigma}]+2\alpha \left\{[R]R^{\S}_{\mu\nu}n^{\mu}n^{\nu}+bK^{\rho}{}_{\rho} +\overline\nabla^{\rho}\overline\nabla_{\rho}[R] \right\},\\
\kappa n^{\nu}h^{\rho}{}_{\mu}[T_{\rho\nu}]&=&(1+2\alpha R_{\S})\overline\nabla^{\rho}[K_{\rho\mu}]\\
&+&2\alpha\left\{[R]n^{\nu}h^{\rho}{}_{\mu}R_{\rho\nu}-\overline\nabla_{\mu} b +\left[K_{\rho\mu}\right]\overline\nabla^{\rho}R_{\Sigma}+K^{\S}_{\rho\mu}\overline\nabla^{\rho}[R]\right\}.
\eean
These were derived in \cite{S} by direct computation of the discontinuity of the energy-momentum tensor via (\ref{fe}) and thus the previous calculation is an independent proof of them. More importantly, the previous calculation shows that the double layer contribution $\underline{t}_{\mu\nu}$ is {\em necessary} to keep  $\underline{T}_{\mu\nu}$ divergence free: {\em without the $\underline{t}_{\mu\nu}$ contribution the energy-momentum content would not be locally ``conserved''}.

Combining the previous two expressions with (\ref{tauexc}--\ref{taue}) and using the Gauss and Codazzi equations for $\Sigma$ on both sides:
$$
R^{\pm}-2R^{\pm}_{\mu\nu}n^\mu n^\nu = {\cal R} -(K^{\pm\rho}{}_\rho)^2+K^{\pm}_{\mu\nu}K^{\pm\mu\nu},\hspace{7mm}
n^{\mu}R^{\pm}_{\mu\rho}h^{\rho}{}_{\nu}=\overline\nabla^{\mu}K^{\pm}_{\mu\nu}-\overline\nabla_{\nu}K^{\pm\rho}{}_{\rho} 
$$
where ${\cal R}$ is the scalar curvature of $(\Sigma,h_{\mu\nu})$,  one derives the field equations for the energy-momentum content of the double layer \cite{S}:
\bea
n^{\nu}h^{\rho}{}_{\mu}[T_{\rho\nu}]&=&-\overline\nabla^{\nu}\tau_{\mu\nu}-K^{\rho}{}_{\rho}\tau_{\mu}-\overline\nabla_{\mu}\tau , \label{divtau} \\
n^{\mu}n^{\nu}[T_{\mu\nu}]-\tau_{\mu\nu}K^{\mu\nu}_{\S}+\overline\nabla^{\mu}\tau_{\mu}&=&2\frac{\alpha}{\kappa} [R]\left(R^{\S}_{\mu\nu}n^{\mu}n^{\nu}+K^{\mu\nu}_{\S}K_{\S\mu\nu} \right) \nonumber \\
&=&\frac{\alpha}{\kappa} [R]\left(R_{\S}-{\cal R} +(K^{\rho}{}_{\rho})^{2}+K^{+}_{\mu\nu}K^{-\mu\nu}\right).\nonumber
\eea
It may be observed that $\tau_{\mu\nu}$ is not divergence-free {\em even} in cases where there is no flux of energy from the bulk across $\S$ (i.e. $n^{\nu}h^{\rho}{}_{\mu}[T_{\rho\nu}]=0$) due to new ``source'' terms $\tau_\mu$ and $\tau$. Note, however, that if one sets $[R]=0$ then the standard case in general (non-quadratic) $F(R)$ theories is recovered, because then $[R]=0$ is a necessary condition \cite{DSS,S}. On the other hand, the case $\alpha=0$ is simply GR ---where a discontinuous $R$ is allowed--- supplemented with (\ref{Kcont}). This implies that in the limit $\alpha \rightarrow 0$ ---or approximately when $\alpha R \ll 1$--- one recovers a GR thin shell {\em but} subject to the condition (\ref{Kcont}), which as remarked above forbids the typical umbilical branes and restricts drastically the possibility of having $\mathbb{Z}_{2}$ symmetric ones. Furthermore, in this limit only the energy-momentum tensor $\tau_{\mu\nu}$ survives {\em with} the additional property that it is trace-free: $\tau^{\mu}{}_{\mu}=0$.

As one can see, $\underline{t}_{\mu\nu}$ is decoupled from the field equations for the rest of the objects with support on $\S$. This seems to allow for the extreme exotic situation in which all $\tau_{\mu\nu}$, $\tau_\mu$ and $\tau$ vanish while $\underline{t}_{\mu\nu}$ does not, describing a shell with only the double-layer energy-momentum contribution. For this to happen, and keeping $[R]\neq 0$ in order to have a non-zero $\underline{t}_{\mu\nu}$, the following are required
$$
K^\rho{}_\rho=0, \hspace{3mm} \overline\nabla_\mu [R]=0, \hspace{3mm} b = n^\mu \left[\nabla_{\mu}R\right] =0, \hspace{2mm} [K_{\mu\nu}] +2\alpha [RK_{\mu\nu}]=0.
$$
The second and third imply that $R^\pm$ are constants on $\S$, while the first one implies that $\S$ must have zero mean curvature. These exceptional cases can be called ``pure double layers".

On the other hand, if $\underline{t}_{\mu\nu}=0$ (which requires $[R]=0$) then the other new terms $\tau_{\mu}$ and $\tau$ vanish too, leading to a standard shell supplemented with (\ref{Kcont}).

In what follows, I present some simple examples of gravitational double layers with explicit formulas for all the quantities involved. The examples are built in constant curvature bulks ---that is, (anti) de Sitter ((A)dS) or flat spacetimes--- with $\S$ separating two such  manifolds with different cosmological constants $\Lambda^+\neq \Lambda^-$. The line-element on each $\pm$ side reads
\be
ds^2_\pm=-(k-\frac{R_\pm}{20} x^2_\pm)dy^2_\pm+(k-\frac{R_\pm}{20} x^2_\pm)^{-1} dx^2_\pm + x^2_\pm d\Omega^{2}_{k} \label{AdS}
\ee
where $k=\pm 1,0$ and $d\Omega^{2}_{k}$ is the complete Riemannian 3-dimensional metric of constant curvature $k$. The constant scalar curvatures $R_\pm$ are related to the cosmological constants by $\Lambda_\pm =R_\pm (3+\alpha R_\pm)/10$. The ranges (and causal character) of the coordinates $\{x_{\pm},y_{\pm}\}$ depend on the signs of the constants $R_{\pm}$ and $k$. We need to find corresponding hypersurfaces in both $\pm$ sides such that the condition (\ref{Kcont}) is satisfied. Restricting to hypersurfaces respecting the spherical symmetry, so that $\S$ is given by means of the parametric expressions $x_{\pm}=x_\pm(t)$ and $y_{\pm}=y_\pm(t)$ on each side, the equality of the inherited first fundamental forms on $\S$ requires
\be
x_+(t) = x_-(t)\equiv a(t) ; \hspace{3mm} \frac{dy_\pm}{dt}=\frac{\varepsilon_\pm}{\A}\sqrt{\B} \label{eqs}
\ee
where $\varepsilon_\pm^2=1$, dots stand for derivatives with respect to $t$ and the first fundamental form on $\S$ is a Robertson-Walker metric
\be
d\gamma^{2}= -dt^{2}+a^{2}(t) d\Omega^{2}_{k}. \label{RW}
\ee
For later reference, I write the GR energy-momentum quantities as they would be computed by scientists living on $\S$ but unaware that this is actually a double layer of a higher dimensional bulk, so that they would use GR as their gravitational theory along the lines explained in more detail in \cite{MSV,MSV1}. These are denoted by $\varrho^{GR}$ (GR energy density), $p^{GR}$ (GR pressure) and $\Lambda^{GR}$ (GR 4-dimensional cosmological constant) and given by 
\be
8\pi G \varrho^{GR} +\Lambda^{GR}= 3\frac{\dot a^{2}+k}{a^{2}}, \hspace{2cm} 8\pi G  (\varrho^{GR}+3p^{GR})-2\Lambda^{GR} =-6\frac{\ddot a}{a} .\label{GR}
\ee

Coming back to the double layer, the unit normals on each side of $\Sigma$ read
\be
\bm{n}_\pm =-\epsilon_\pm (-\dot a dy_\pm +\dot y_\pm dx_\pm) \label{n}
\ee
where the signs $\epsilon_\pm$ determine the part of the $\pm$-side of the bulk to be matched.
Then, the second fundamental forms on each side can be computed to be
\be
K^\pm_{\mu\nu}dx^\mu dx^\nu = (\varepsilon\epsilon)_\pm \left(-\frac{\ddot a -\frac{R_\pm}{20}a}{\sqrt{\B}}\, dt^2 +a\sqrt{\B}\, d\Omega^{2}_{k} \right) \label{K}
\ee
so that the indispensable condition (\ref{Kcont}) leads to two possible solutions for the scale factor $a(t)$ depending on whether or not $\dot a=0$.

{\bf Case (i):} $a$ is a constant given by the relation
$$
16(\Aplus)(\Aminus) = 1
$$
with both factors $\A>0$, together with $k=(\epsilon\varepsilon)_+/(\epsilon\varepsilon)_-$. This forbids $k=0$ in this case, so that flat space-time cannot be described by $\S$. Observe that one side of the bulk, however, can certainly describe 5-dimensional flat space-time, by choosing $R_-=0$, say, and then $a^2 =\frac{75}{4} \frac{1}{R_+} >0$ and $k=1$. All cases with $k=1$ have a line-element on the layer corresponding to the Einstein static universe. The non-zero eigenvalues of $\tau_{\mu\nu}$ are easily computed from (\ref{tauexc1}), where now $b=0$, to give the energy density and pressure of the double layer
\bean
\kappa \varrho = (\epsilon\varepsilon)_-  \frac{R_-}{5}a(1+2\alpha R_-)\sqrt{\Aplus} -
(\epsilon\varepsilon)_+  \frac{R_+}{5}a(1+2\alpha R_+)\sqrt{\Aminus} ,\\
\kappa p = \frac{(\epsilon\varepsilon)_-}{a}(1+2\alpha R_-) \sqrt{\Aminus}-
\frac{(\epsilon\varepsilon)_+}{a}(1+2\alpha R_+) \sqrt{\Aplus}
\eean
which are constant but different from the GR values. Concerning the other quantities on the double layer one has from (\ref{tauex},\ref{taue}) 
$$
\tau_\mu =0, \hspace{1cm}  \kappa \tau = 2\alpha [R] (\epsilon\varepsilon)_\pm\left(\frac{3}{a}\sqrt{\A} -\frac{R_\pm a}{20\sqrt{\A}}\right)
$$
where any of the two signs is valid, and finally formula (\ref{t}) holds as such.

{\bf Case (ii):} $\dot a \neq 0$, then $a(t)$ is the solution of (recall that $R_\S=(R^++R^-)/2$)
\be
\dot a^2 +k = \frac{a^{10}}{c^2}\left(\frac{[R]}{40}\right)^2+a^2 \frac{R_\S}{20}+\frac{c^2}{4a^6} \label{a}
\ee
where $c\neq 0$ is an arbitrary constant. This can also be written in any of the two alternative forms
\be
\B =\left(\frac{c}{2a^3}\mp\frac{a^5}{40c}[R] \right)^2 . \label{sol}
\ee
By analyzing equation (\ref{a}) in the standard ``kinetic + potential energy" form $\dot a^2 +V(a)=0$, one gets the qualitative behaviour of the solutions $a(t)$. The ``potential'' $V(a)$ is an even function. In the physical region with $a>0$, $V(a)$ has a unique maximum and its value is always negative for the cases with $k=0,-1$. Therefore, in these cases the double layers describe a 4-dimensional Universe starting from a big-bang followed by a decelerated expansion phase until $\dot a$ reaches a minimum value from where the Universe undergoes an accelerated expansion epoch leading to unbounded values for $a$ and $\dot a$. For the remaining, closed, case with $k=1$, the maximum of $V(a)$ may be positive or negative depending on whether $|c|$ is small or large, respectively. In the latter case the solutions behave just as in the previous cases with $k=0,-1$. In the former case, on the contrary, the possible solutions for $a(t)$ have two branches. In one of them the double layer describes a Universe which, from a big-bang, reaches a re-collapsing time and then contracts to a big crunch. The other branch, however, is singularity-free, starting with a large Universe that contracts to a minimum volume and then re-expands with accelerated expansion. Of course, there is a critical value of $|c|$ between the two mentioned possibilities, where the Universe tends asymptotically to an Einstein static solution of the type given in case (i).

The hypersurface $\S$ describing the double layer can be given explicitly as a submanifold of (A)dS (or flat) space-time by integrating 
(\ref{eqs}). For instance, for $k=0$ one gets
$$
y_{\pm }-y_{0\pm}=\frac{5\varepsilon_{\pm}}{2R_{\pm}}\left(\frac{[R]}{20c}x_{\pm}^{4} \pm \frac{c}{x_{\pm}^{4}}\right)
$$
and more complicated (but explicit in terms of elementary functions) formulae in the other cases $k=\pm 1$. Using again that $b=0$, the non-zero eigenvalues of $\tau_{\mu\nu}$ are easily computed from (\ref{tauexc1}) 
\bea
\kappa \varrho = -(\epsilon\varepsilon)_{+}\left\{\frac{3c}{a^{4}} \left(1+\alpha(R_{+} +R_{-})\right)+\alpha \frac{a^{4}}{4c} [R]^{2} \right\} ,\label{energy}\\
\kappa p = -(\epsilon\varepsilon)_{+}\left\{\frac{c}{a^{4}} \left(1+\alpha(R_{+} +R_{-})\right)-\alpha \frac{a^{4}}{20c} [R]^{2} \right\} \label{pressure}
\eea
which give the energy density and pressure within the double layer. Observe that this has two components, one dominates for small values of $a$, the other for large values of $a$. The former has an equation of state of radiation type $p=\varrho/3$, and is the only one surviving for the GR limit when $\alpha \rightarrow 0$. The latter has a strange equation of state $p=-\varrho/5$ and is proportional to the square of the difference between the constant curvatures at both sides of the double layer. Observe that cases with de Sitter on one side and anti de Sitter on the other side are feasible, as the signs of $R_{\pm}$ are free. There are cases with flat space-time on one side too.

One also has from (\ref{tauex},\ref{taue}) 
$$
\tau_\mu =0, \hspace{1cm}  \kappa \tau = - (\epsilon\varepsilon)_+
\alpha \frac{2a^{4}}{5c}[R]^{2}
$$
while formula (\ref{t}) holds as such (and is independent of the signs $\epsilon_{\pm}$ and $\varepsilon_{\pm}$). Observe that (\ref{energy}) and (\ref{pressure}) satisfy the following generalized continuity equation
$$
\dot \varrho +3\frac{\dot a}{a} (\varrho +p) =\dot \tau 
$$
which is a particular case of the general expression (\ref{divtau}) and reflects the novel fact that the energy-momentum $\tau_{\mu\nu}$ of the double layer is not divergence free in general. Therefore, the traditional behavior of $\varrho\sim a^{-3(1+w)}$ for an equation of state $p=w\varrho$ does not hold for double layers in general. This opens the door for new physical behaviors.

Expressions (\ref{energy}) and (\ref{pressure}) are to be compared with the GR values (\ref{GR}) as they would be computed by scientists on $\S$ unaware of its double layer character. These are
\bean
8\pi G \varrho^{GR} = \frac{3}{4}\left( \frac{a^8}{c^2} \frac{[R]^2}{400}+\frac{c^2}{a^8}\right), \hspace{1cm}
8\pi G p^{GR} = \frac{1}{4}\left( \frac{-11a^8}{c^2} \frac{[R]^2}{400}+\frac{5c^2}{a^8}\right),\\
 \Lambda^{GR}=\frac{3}{40} \left( R_+ + R_-\right). \hspace{4cm}
 \eean
Note that the GR cosmological constant would be simply proportional to the sum of the constant scalar curvatures at both sides of the double layer. Hence, in order to have a positive $ \Lambda^{GR}$, at least one side of the bulk should have a positive constant curvature. 

\section*{Acknowledgements}
I thank the Center for Astrophysics at Shanghai Normal University, and especially Prof. Roh-Suan Tung, for hospitality. Comments from Prof. L. Bel  and Dr. J. Camps are acknowledged. This work was supported by the National Natural Science Foundation of China under grant No. 11071167. The author is also supported by grants
FIS2010-15492 (MICINN), GIU12/15 (Gobierno Vasco), P09-FQM-4496 (J. Andaluc\'{\i}a---FEDER) and UFI 11/55 (UPV/EHU).

\end{document}